\documentclass[12pt]{article}
\usepackage[english]{babel}
\usepackage[cp1250]{inputenc}

\usepackage{setspace}
\usepackage[dvips]{graphicx}
\usepackage{amssymb}

\usepackage[dvips]{graphics}

\begin{document}

\title {Kullback-Leibler quantum divergence as an indicator of quantum chaos}

\author{A. Kowalewska-Kud{\l}aszyk$^{1}$, J. K. Kalaga$^{2}$, W. Leo\'nski\footnote{Corresponding author - email: wleonski@proton.if.uz.zgora.pl} $^{2}$ and V. Cao Long$^{2}$}

\date{}

\maketitle

\begin{center}
%
%
{\small\it $^1$Nonlinear Optics Division, Department of Physics, Adam Mickiewicz University,\\ Umultowska 85, 61-614 Pozna\'n, Poland}

\vskip 0.2cm
%
%
{\small\it $^2$Quantum Optics and Engineering Division, Institute of Physics,  University of Zielona G\'ora,\\ ul.~Prof.~A.~Szafrana~4a, 65-516 Zielona G\'ora, Poland}

\end{center}

\vskip 0.5cm

\begin{abstract}
We discuss a system of a nonlinear Kerr-like oscillator externally pumped by ultra-short, external, coherent pulses. For such a system, we analyse the application of the Kullback-Leibler quantum divergence $K[\rho||\sigma]$ to the detection of quantum chaotic behaviour. Defining linear and nonlinear quantum divergences, and calculating their power spectra, we show that these parameters are more suitable indicators of quantum chaos than the fidelity commonly discussed in the literature, and are useful for dealing with short time series. Moreover, the nonlinear divergence is more sensitive to chaotic bands and to boundaries of chaotic regions, compared to its linear counterpart.
\end{abstract}

\section{Introduction}
Investigations of the problems of chaos in the dynamics of a quantum system are of current interest. These problems are particularly important whenever quantum systems are used in quantum information theory models, especially in the context of the problem of decoherence processes in the practical realization of quantum computations. Processes of this type induced by interactions with the environment or by any imperfections present in a quantum computer \cite{Z03,FFS04}, lower the efficiency of entanglement creation between quantum states. Therefore, one of the problems concerning chaotic dynamics is to determine whether a system exhibits regular or chaotic evolution. For classical systems, the methods for the determination of the boundaries between regular and chaotic dynamics have already been developed and are widely used (Lyapunov exponents, power spectra or entropy \cite{SJ05}), whereas in quantum dynamics new methods must be found. Because of the linearity of the Schr{\"{o}}dinger equation, the methods used in classical dynamics cannot be used in quantum theory.
Therefore, in order to identify features that are not related to quantum chaotic behaviour, it is important to consider quantum systems which exhibit chaotic behaviour in their classical limit.
In particular, certain methods have already been mentioned in \cite{H92} and are currently discussed in the papers devoted to the problems of quantum chaotic evolution. For instance, 
the theory of random matrices can predict the statistical properties of energy level fluctuations \cite{BFFMPW81} and there is an evident correspondence between the successive eigenstates of quantum chaotic systems and the eigenstates of random matrices \cite{BGS84,BT77,I87,KMH88,HZ90}.
There is also a method based on the decay of the fidelity between two quantum states (unperturbed and perturbed) \cite{P84,EWLC02,WLT02}. It has been proven that in a region where a quantum system behaves chaotically, fidelity decays exponentially \cite{JP01,JSB01,PZ02}. Moreover,  the entanglement in the evolving system grows when the dynamics of a quantum systems reaches the regions of chaotic behaviour \cite{WGSH04}.

In this paper, we propose to apply parameters based on the Kullback-Leibler divergence (KLD) as indicators of quantum chaotic behaviour. By discussing a system of a pumped nonlinear Kerr-like oscillator, we show that these parameters are not only indicators of quantum chaotic dynamics of the system but they also behave chaotically in the classical sense, despite their quantum nature. In particular, we discuss the KLD and show that the linear and first nonlinear terms in its expansion can clearly show the boundaries between regular and chaotic motion in the dynamics of a quantum system.  

\section{Quantum divergence}
Even though it is not a true distance measure, the quantum Kullback-Leibler divergence $K_{KL}$ (also termed relative entropy) is widely used in probability theory and information theory for comparisons between two probability distributions.
Thus, for the continuous random variables $P$ and $Q$ described by their densities $p(x)$ and $q(x)$, the relative entropy is defined as \cite{KL51,K59}:
\begin{equation}
K_{KL}\left[P||Q\right]=\int\limits_{-\infty}^{\infty}p(x)\log\frac{p(x)}{q(x)}dx\,\,\, .
\label{eq1}
\end{equation}
The divergence $K_{KL}$ is well-defined for continuous variables and it is always non-negative ($K_{KL}\left[P||Q\right]\geq 0$), asymmetric ($K_{KL}\left[P||Q\right]\neq K_{KL}\left[Q||P\right]$) and invariant under the transformation of parameters.

In quantum information theory, for the comparison of two density matrices $\rho$ and $\sigma$ describing two quantum states of a system, the quantum counterpart of the Kullback-Leibler divergence $K_{KL}$ is used and defined as follows \cite{U62}:
\begin{equation}
K_{KL}\left[\rho||\sigma\right]=Tr\left[\rho\left(\ln\rho-\ln\sigma\right)\right]\,\,\, .
\label{eq2}
\end{equation}
From the definition (\ref{eq2}), it follows that the value of quantum divergence is zero only for $\rho=\sigma$, while positive values measure the extent of difference between the two density matrices and, consequently, between the two quantum states. It is also known that this quantum counterpart of the Kullback-Leibler divergence, eq.(\ref{eq2}), 
is singular whenever the reference state is a pure state \cite{A03}. In order to circumvent this problem, one can introduce the quantum $q$-divergence defined as \cite{A03}:
\begin{equation}
K_q\left[\rho||\sigma\right]=D_{q}Tr\left(\rho^x\sigma^{1-x}\right)|_{x\rightarrow 1-0}\,\, ,
\label{eq3}
\end{equation} 
where $D_q$ is Jackson’s differential operator, which can be expressed as:
\begin{equation}
D_{q}f(x)=\frac{f(qx)-f(x)}{x(q-1)}\,\, ,
\label{eq3a}
\end{equation}
satisfying the identity:
\begin{equation}
D_q\left(f(x)g(x)\right)=\left(D_qf(x)\right)g(x)+f(x)\left(D_qg(x)\right)+x(q-1)\left(D_qf(x)\right)\left(D_qg(x)\right)\,\, .
\label{eq3b}
\end{equation}
In the limit $q\rightarrow 1$, the quantum $q$-divergence tends to the quantum divergence defined in (\ref{eq2}).
The quantum q-divergence can be applied to determine the purity of the states, which is one of the most important problems in quantum information theory.
 
The importance of using the Kullback-Leibler divergence in information theory results from the fact that many other quantities 
can be interpreted as the results of applying $K_{KL}$ to specific cases. 
For instance, mutual information can be expressed by means of the Kullback-Leibler divergence as given by \cite{CT06}:
\begin{equation}
I(X;Y)=K_{KL}\left(P(X,Y)||P(X)P(Y)\right)\,\, ,
\label{eq4}
\end{equation}
therefore, it is the divergence between the product of two probability distributions $P(X)$ and $P(Y)$ and the joint probability distribution $P(X,Y)$. 
Moreover, the Kullback-Leibler divergence is related to the Shannon entropy via the relation:
\begin{equation}
H(X)=\log{N}-K_{KL}\left(P(X)||P_{U}(X)\right)\,\, ,
\label{eq5}
\end{equation}
where $P(X)$ is the true probability distribution and $P_{U}(X)$ is the uniform probability distribution.

In this paper, we apply the quantum divergence (\ref{eq2}) as a tool for the analysis of quantum dynamics of the system allowing us to determine whether or not the chaotic regions have been reached.
Several methods are used to determine whether a quantum system exhibits chaotic behaviour. One of them is based on the fidelity between two quantum states described by wave-functions and has been discussed in the literature \cite{P84,WLT02,EWLC02}. One of these states is generated by the standard mapping procedure, whereas the other evolves under a slightly perturbed evolution operator. It has been shown that whenever the system behaves chaotically (in the quantum chaotic sense), the fidelity between theses states decays exponentially. Since the quantum divergence (\ref{eq2}) is defined with respect to two density matrices, in this paper we use density matrices to describe the evolution of the quantum system. One of the matrices corresponds to the evolution of the unperturbed system, whereas  the other describes the evolution of a slightly perturbed system under study (similarly to fidelity analysis).

In \cite{WLT02,EWLC02} it has been shown that whenever the dynamics of a system exhibits quantum chaotic behaviour, the differences between the analysed states are clearly visible in fidelity evolution. Therefore, we presume that the analysis of quantum divergence would also demonstrate these differences, and consequently, that quantum divergence could also be used  to determine the boundary between regular and chaotic behaviour in quantum systems.

\section{The model}
For the purpose of analysis of the quantum divergence (\ref{eq2}) in quantum chaos theory, it would be convenient to use a system whose ability to
demonstrate quantum chaotic behaviour was confirmed by other well-known indicators of quantum chaos. 
In this paper, we consider a system composed of a Kerr-like nonlinear oscillator externally pumped by a series of ultra-short coherent pulses. Systems based on Kerr-like nonlinearity are commonly used as models for quantum-optical investigations, for a comprehensive review see for instance \cite{PL94,T03} (\textit{and references cited therein}). Such systems can be a source of various quantum states of the electromagnetic field. For instance, Miranowicz \textit{et al.} \cite{MTK90} were the first to show that the discrete superpositions of arbitrary numbers of coherent states (Schr\"odinger cats or kitten states) can be generated by systems involving Kerr-like nonlinearities. These states are one of the most commonly discussed states of quantum optics. In general,  models with nonlinear media have been widely studied in the context of both nonlinear and quantum optics, and have been described in various review papers, e.g. \cite{PP00,BDF01}.

It is known that the classical counterpart of systems based on Kerr-like nonlinearity can demonstrate regular or chaotic behaviour, depending on their characteristic parameters\cite{SG01,SMG03,SGM06}.  Particularly, the influence of the strength of the external pumping of the nonlinear oscillator on the nature of the dynamics of the system was discussed in \cite{L96}, for cases with or without damping.

The system considered here contains a nonlinear Kerr-like oscillator and is externally pumped by a series of ultra-short coherent pulses. It is described by the following Hamiltonian:
\begin{equation}
\hat{H}=\hat{H}_{NL}+\hat{H}_{K}\,\,,
\label{eq5a}
\end{equation}
where $\hat{H}_{NL}$ represents the nonlinear Kerr-like oscillator, whereas $\hat{H}_K$ describes the interactions with external pumping. In particular, $\hat{H}_{NL}$ and $\hat{H}_K$ can be expressed as (in units of $\hbar =1$):
\begin{eqnarray}
\hat{H}_{NL}&=&\frac{\chi}{2}\left(\hat{a}^\dagger\right)^2\hat{a}^2\,\,,\label{eq6a}\\
\hat{H}_{K}&=&\epsilon\left(\hat{a}^\dagger+\hat{a}\right)\sum\limits_{k=1}^\infty\delta(t-kT)\,\,,
\label{eq6b}
\end{eqnarray}
where $\chi$ is the nonlinearity constant, $\epsilon$ characterizes the strength of the interaction between the nonlinear oscillator and the external filed, $\hat{a}^\dagger$ and $\hat{a}$ are the usual boson creation
and  annihilation operators, respectively. The parameter $T$ denotes the time interval between two subsequent external pulses. Here, we assume that $T$ significantly exceeds the inverse of the external field frequency. Hence, the series of the ultra-short external pulses can be modelled by a sum of Dirac-delta functions.

For simplicity, our model  neglects damping and, in consequence, we can use the wave-function approach in order to describe the dynamics of the system. In particular, in order to determine the dynamics of the system, we construct the unitary evolution operators on the basis of the Hamiltonians (\ref{eq6a},\ref{eq6b}) and apply them repeatedly to the initial state of the system. Consequently, we obtain a quantum map describing the evolution of our system:
\begin{equation}
|\Psi_u(n)\rangle=\left(\hat{U}_{NL}\hat{U}_K\right)^n|\Psi(t=0)\rangle\,\, .
\label{eq7}
\end{equation}
The operator $\hat{U}_{NL}$ describes the unitary evolution of the system between two subsequent external pulses (corresponding to a time interval T), according to the formula: 
\begin{equation}
\hat{U}_{NL}=e^{-i\chi T\hat{n}(\hat{n}-1)}\,\, ,
\label{eq8}
\end{equation}
whereas the operator $\hat{U}_K$ describes the interaction between the Kerr-like oscillator and a single ultra-short pulse. This operator can be written as:
\begin{equation}
\hat{U}_{K}=e^{-i\epsilon\left(\hat{a}^\dagger+\hat{a}\right)}\,\, .
\label{eq8a}
\end{equation}
In fact, this method has been applied to the investigations of  finite-dimensional quantum state generation \cite{LM01} and to the comparison of the features of quantum and classical dynamics in  nonlinear kicked systems  \cite{L96}. At this point one should mention another, very interesting method that allows to find the solution of the problems related to the nonlinear oscillators' models discussed in \cite{BDGN2007}.

In particular, as already shown in \cite{L96}, the classical counterpart of the system considered here can exhibit both regular and chaotic dynamics. For fixed values of the nonlinearity parameter $\chi$ and the time interval between two subsequent pulses $T$ and by varying the strength of interaction between the system and the external field, one can obtain regular behaviour of the system (if the interactions are weak) or chaotic behaviour in the classical sense (if the excitation is strong). The values of the parameters leading to a given character of the system's behaviour can be read off from the bifurcation diagram. Thus, using the method described in \cite{L96}, we generated the diagram (Fig.1) showing the nature of the dynamics of the system depending on the value of the parameter $\epsilon$. Fig.1 shows the values of the real and imaginary parts of the complex parameter  $\alpha$ which is the classical counterpart of the annihilation operator $\hat{a}$. $|\alpha |^2$ is the energy of the system that is the classical counterpart of the considered system.  If the value of $\epsilon$ increases from $\sim 0.3$ to $\sim 0.7$, we observe a region of regular dynamics followed by a chaotic band, regular window and a region of deep chaos, successively.

As mentioned in \cite{KKL08,KKL09}, the considered system exhibits quantum chaotic behaviour and, therefore, it constitutes a good model for testing the usefulness of various parameters for detecting chaotic regions and regions preceding the quantum chaotic regions. 
In this paper, we intend to find out whether the quantum divergence is sensitive to quantum chaos and whether its behaviour changes near the chaos boundary. 
As mentioned above, apart from the regions of clearly regular dynamics and a region of purely chaotic dynamics (a case of ``deep chaos''), the classical pumped nonlinear oscillator exhibited a small chaotic region (chaotic band) for $\epsilon\sim 0.36$. As shown in \cite{KKL09}, in this region, the fidelity did not indicate quantum chaos in the dynamics of the system.
Here, we examine whether or not the quantum divergence is sensitive to such chaotic bands.

\section{Analysis of quantum divergence}
Since the definitions (eqns.(\ref{eq2},\ref{eq3})) of the KLD involve density matrices describing the evolution of the system, we applied quantum mapping to generate density matrices, subsequently used to determine quantum KLD. 

To determine the quantum divergence, we need density matrices for the evolution of unperturbed and perturbed systems. While the unperturbed density matrix can be derived from the wave-function generated by means of the quantum mapping (eq.(\ref{eq7})), its perturbed counterpart can be generated by a similar procedure where we replace the unitary kick operator (eq.(\ref{eq8a})) by another operator $\hat{U}_{Kp}$, defined by the relation:
\begin{equation}
\hat{U}_{Kp}=e^{-i\left(\epsilon+
\Delta_{\epsilon}\right)\left(\hat{a}^{+}+\hat{a}\right)}\,\,,
\label{eq9b}
\end{equation} 
where $\Delta_{\epsilon}$ describes a small perturbation in the laser field strength. In further considerations we assume  $\Delta_{\epsilon}=0.001$.
Therefore, the perturbed density matrix can be expressed as:
\begin{equation}
\rho_p(n)=
\left(\hat{U}_{NL}\hat{U}_{Kp}\right)^n|\Psi(t=0)\rangle
\langle\Psi(t=0)|\left(\hat{U}^{+}_{Kp}\hat{U}^{+}_{NL}\right)^n\,\, .
\label{eq8b}
\end{equation}

The quantum divergence defined by (\ref{eq2}) is singular when the reference state is a pure quantum state. We deal with a pumped oscillator without damping and for this reason, instead of using the exact definition (\ref{eq2}) we shall apply the first terms in the series expansion of the logarithm function and thus use the relation \cite{AS64}:
\begin{equation}
\ln\rho-\ln\sigma=\sum\limits_{n=1}^{\infty}\left(-1\right)^{n+1}\frac{\left(\rho-1\right)^{n}}{n}-\sum\limits_{n=1}^{\infty}\left(-1\right)^{n+1}\frac{\left(\sigma-1\right)^{n}}{n}\,\, .
\label{eq9}
\end{equation}
In particular, we shall concentrate on the first and second terms of such expansion in order 
to investigate their usefulness for the analysis of  quantum chaotic systems.

Thus, we first use the terms corresponding to $n=1$ (linear terms) and from the definition (\ref{eq9}) it follows that we obtain a \textit{linear} quantum divergence (LQD) in the form:
\begin{equation}
K^{(1)}_{KL}=Tr\left[\rho(\rho-\sigma)\right]\,\,,
\label{eq10}
\end{equation}
If we take the first two terms of the expansion (\ref{eq9}) we obtain
\begin{equation}
K^{(2)}_{KL}=Tr\left[\rho\left(\rho-\frac{1}{2}(\rho-1)^2-\sigma+\frac{1}{2}(\sigma-1)^2\right)\right]\,\,.
\label{eq11}
\end{equation}
The divergence thus defined contains both the linear and the first nonlinear term. In further considerations, $K^{(2)}$ will be referred to as the \textit{nonlinear} quantum divergence (NQD).

\subsection{Linear quantum divergence}
As proved in \cite{WGSH04}, the von Neumann entropy (indicating the existence of bipartite entanglement) increased rapidly when the initial state of the system exhibited the behaviour characteristic of quantum chaos followed by irregular changes (of small amplitude) depending on the number of qubits involved in the system. 
In our considerations we shall utilise the LQD and NQD descriptors in order to determine whether the system is chaotic. 

First, we concentrate on the LQD parameter $K^{(1)}_{KL}$. For regular motion, the function $K^{(1)}_{KL}$ oscillates regularly (between $0$ and $1$) in time (described by the number of pulses) (Fig. 2). This means that small perturbations in the laser excitation strength ($\Delta_\epsilon$) do not influence the final state of the pumped oscillator and, in fact, the perturbed and unperturbed states are the same quantum states. The same type of oscillations was observed while considering the fidelity between the two quantum states discussed in \cite{KKL09}. 

Moreover, we observed the behaviour analogous to that discussed in \cite{KKL09}, when the system was  inside the regular window and the value of $\epsilon$ was close to that corresponding to the  boundary of the region of deep chaos ($\epsilon=0.46$). Fig.3 shows that $K^{(1)}_{KL}$ oscillates and these oscillations are modulated by sine-like variations -- the nature of these changes in $K^{(1)}_{KL}$ differs from that depicted in Fig.2. In the long-time limit, $K^{(1)}_{KL}$  $\sim 1$ was reached after about $6\times 10^4$ external pulses. This proves the importance of  analysis in the long-time limit. Moreover, the effect of the observed modulations constitutes an example of quantum beats generated in a system with Kerr-like nonlinearity. Therefore, when the quantum system was close to the boundary of deep chaos, even in the region of regular dynamics, the changes in $K^{(1)}_{KL}$ were no longer of the regular oscillatory nature (characteristic of the regular motion of a quantum system) and quantum beats appeared instead.

Furthermore, we analysed the changes in $K^{(1)}_{KL}$ in the frequency domain. This was particularly useful when the final signal was a composition of several oscillations with different amplitudes. Whenever the oscillations had a single frequency (as for the regions of regular dynamics -- Fig.2),  plotting the power spectra was unnecessary, as opposed to the regions of chaotic dynamics.
 
The power spectra of the LQD and NQD can be obtained by calculating the Fourier transform of $K^{(1)}_{KL}$ (or $K^{(2)}_{KL}$) and the square of its absolute value and subsequently normalising it. 

Fig.4 shows the time-dependence of the LQD (Fig.4a) and its power spectrum (Fig.4b) for the system situated deep inside the chaotic region ($\epsilon =0.7$). $K^{(1)}_{KL}$ increased with the first $\sim 2\times 10^3$ pulses and then exhibited chaotic changes close to unity (Fig. 4 a). The time dependence of the linear part of the entropy consisted of several oscillations of different frequencies (Fig. 4b). We assume that these frequencies were
distributed regularly in the power spectrum, since several groups of these frequencies were observed.
The combination of these oscillations resulted in irregular changes in $K^{(1)}_{KL}$ (Fig.4a).  
Therefore, the exponential decay of the fidelity and its further irregular changes of small amplitude \cite{KKL09}, as well as the initial monotonic increase and further irregular oscillations in the LQD are characteristic features of quantum chaos. These irregularities were composed of several groups of oscillations of various frequencies (see the power spectrum).

\subsection{Nonlinear quantum divergence}
Since the chaotic behaviour is the domain of nonlinear systems, it can be useful to introduce the nonlinear term in the expansion of quantum divergence in our considerations. Fig. 5a shows the time dependence of the NQD parameter $K^{(2)}_{KL}$ for weak external excitations. This situation corresponds to regular dynamics and we observe slow oscillations of $K^{(2)}_{KL}$ modulated by other oscillations of high frequency. In Fig.5b we see that only a few frequencies were involved in the dynamics of the system and we can identify them in the power spectrum. The slowly varying changes in $K^{(2)}_{KL}$ were of the same frequency as those for the LQD $K^{(1)}_{KL}$ (Fig. 2). The addition of the first nonlinear term introduced fast oscillations.

For the first chaotic region (chaotic band between two regular regions -- $\epsilon\approx 0.36$), the linear part of quantum divergence exhibited almost identical behaviour as for $\epsilon =0.1$, just like the fidelity discussed in \cite{KKL09}. However, the addition of the first nonlinear term to the quantum divergence led to visible changes in its behaviour. Although, in general, the character of the oscillations was preserved, fast oscillations were more pronounced than for $\epsilon =0.1$ (Fig. 6a). Much more pronounced changes were observed in the power spectrum. In contrast to the cases already discussed, instead of single and sparse lines  (Fig.5b), a multi-peak structure spreading over the whole spectrum was observed. The time dependence of the nonlinear part of quantum divergence was determined by the sum of a large number of oscillations with various frequencies. This is typical of chaotic motion.
Moreover, by comparing these results with those obtained for the LQD $K^{(1)}_{KL}$ (Figs. 2-4), we observed that the results for the NQD differ considerably from those for the LQD.
Neither the fidelity, nor the LQD were sensitive to this region of chaos (which is visible in the bifurcation diagram), suggesting that this region of chaos appears in the classical dynamics only. On the other hand, a strictly quantum parameter (such as quantum divergence and, consequently, its nonlinear terms) was able to detect the changes in the dynamics of the system in this region. Therefore, the NQD seems to be more sensitive to chaotic dynamics than the other parameters mentioned here. 

For $\epsilon=0.46$, corresponding to the regular window in the bifurcation diagram, the LQD parameter $K^{(1)}_{KL}$ exhibited changes indicating the similarity with quantum chaotic dynamics in the long-time limit; quantum beats appeared (Fig. 3). This was a long-time effect. The NQD, even in the short-time limit, exhibited significant changes in the character of the oscillations, as compared with those for the smaller values of $\epsilon$.  No vanishing oscillations of  $K^{(2)}_{KL}$ occurred (Fig. 7). The changes in the value of  $K^{(2)}_{KL}$ were more complicated and cannot be described by any regular functions (Fig.7a), particularly when looking at the power spectrum of  $K^{(2)}_{KL}$ (Fig.7b). Again,
the final shape of the described function is composed of many oscillations
with different frequencies, however, they are of different character than those described for the linear and nonlinear divergence in the case of the first chaotic band. We observed many frequencies, however, a certain pattern of grouping was clearly visible.

For large values of $\epsilon$ (deep chaos), the nature of the time dependence of  $K^{(2)}_{KL}$ was completely different – no regularities were observed (Fig. 8 a). Moreover, the power spectrum (Fig.8a) was similar to that depicted
in Fig.3a. The frequencies of the oscillations constituting the final time dependence of $K^{(2)}_{KL}$ were concentrated around several values, however, we observed many more groups of frequencies (frequency bands) and the bands were considerably broader than in the spectrum derived from the linear divergence.

Therefore, the addition of the nonlinear term to the quantum divergence allows us to infer additional information about the dynamics of the system. The NQD constructed in this fashion was more sensitive to the chaotic evolution (identified in the bifurcation diagrams for the classical system), especially in regions where the fidelity and LQD indicated no quantum chaos.

The fidelity and LQD indicated quantum
chaotic behaviour in the long-time limit after about $10^{4}$ external pulses. If we were to introduce
damping processes to our model, this time would be too long to observe these
features. This does not necessarily imply that quantum chaos cannot occur in damped systems.  From the analysis of the pumped nonlinear oscillator, in which damping
processes were included (shown in \cite{L96}), it follows that even after $\sim 50$ external
pulses, no further changes in the dynamics of the system occurred and the system was energetically converged. In this case we would not expect the fidelity to reveal the presence quantum chaos. As shown in \cite{KKL09}, if the value of the perturbation $\Delta_\epsilon$ is very small, the character of fidelity decay cannot be used to determine whether or not the system behaves chaotically.
Therefore, one must seek other signatures of quantum chaos in the dynamics of the system, ones that would be sensitive to changes even in such cases. The NLQ parameter $K^{(2)}_{KL}$ obtained by the addition of the nonlinear part to the quantum divergence seems to be appropriate here. Since the changes in the NQD and its power spectra were visible even when the fidelity did not indicate any reaction (and the chaotic behaviour seemed to occur in the classical sense only), we suspect that the NQD can be applied as an indicator of irregular changes in the dynamics of a quantum system for shorter times than those discussed for the fidelity-based indicators.

\section{Summary}
We analysed the application of the Kullback-Leibler quantum divergence $K[\rho||\sigma]$ to the detection of chaotic behaviour in a quantum system. 
In particular, we discussed a system consisting of a nonlinear Kerr-like oscillator externally pumped by a series of ultra-short pulses of coherent field. 

Quantum divergence measures the distance between two probability distributions, in our case -- the differences between two density matrices, one of which was responsible for the evolution of the unperturbed system, whereas the other was defined for a  perturbed system. We showed that these divergence-based parameters can be used to solve the problems of the detection of quantum chaotic behaviours in quantum systems. It is known that the signatures of quantum chaos are still being developed, however, we believe that the quantum K-L divergence can become one of them. 
In particular, we concentrated on the dynamics of the system without damping and used the first one and the first two terms in the expansion of the K-L divergence
in order to avoid the singularity that occurs for pure states. In the system under study, the linear part $K^{(1)}_{KL}$ of the expansion of the K-L divergence (the linear entropy-like parameter) clearly indicated the chaotic behaviour of the system. However, $K^{(1)}_{KL}$ was not sensitive to narrow chaotic bands and to the boundary of the deep chaos region. This behaviour resembles that discussed in \cite{KKL09} for the \textit{fidelity}. 

If the parameter involving the nonlinear term of the expansion (what we termed \textit{nonlinear quantum divergence}, $K^{(2)}_{KL}$) is considered, the situation changes considerably. The power spectrum of $K^{(2)}_{KL}$ serves as an indicator of chaotic behaviour in the evolution of a system even for chaotic bands and when the system is close to the border of deep chaos. This was not the case with the linear parameter $K^{(1)}_{KL}$. The changes in the behaviour of $K^{(2)}_{KL}$ were visible when other parameters (\textit{fidelity} and $K^{(1)}_{KL}$) did not indicate chaotic behaviour, despite the fact that from the bifurcation diagram, the considered regions  could be identified  as chaotic in the classical sense.
In contrast to \textit{fidelity}, the parameters described here can be useful whenever a short-time analysis is required (for instance, for discussions of damped systems). Therefore, we believe that the parameters defined here (and NQD $K^{(2)}_{KL}$ in particular) and their potential applications pave the way towards the description of quantum chaotic processes.

\section{Acknowledgements}
J.K.K. and W.L. acknowledge support from the National Science
Centre under Grant No. N N202 195240.

\newpage
\begin{figure}[p]
\hspace*{1cm}\vspace*{2cm}
\resizebox{14cm}{10cm}
                {\includegraphics{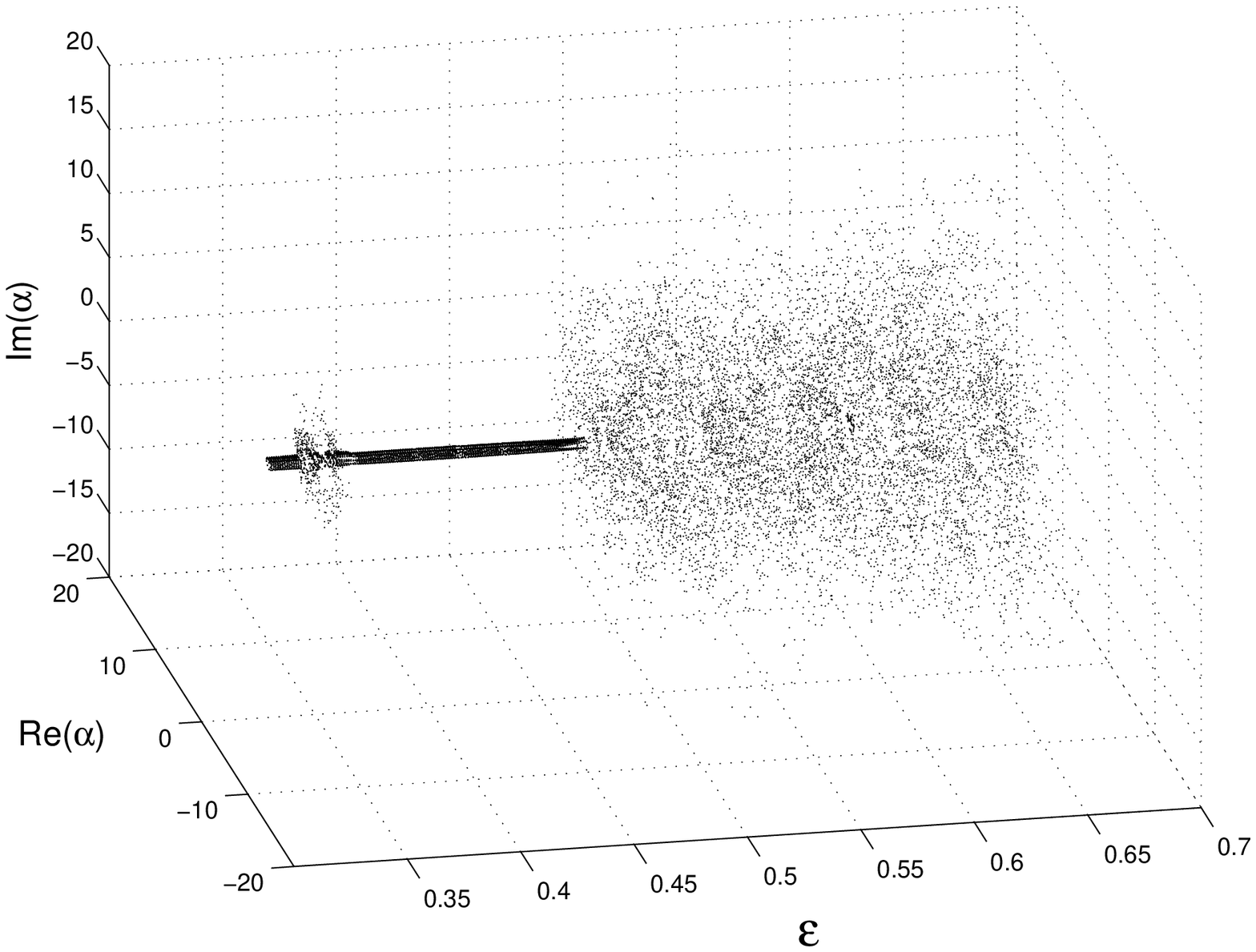}}
\caption{Bifurcation diagram (3D) for the classical counterpart of the discussed model as a function of the strength of external excitation $\epsilon$. 
The parameters are: $T=\pi$, $\chi=1$.}
\end{figure}
\newpage

\begin{figure}[p]
\hspace*{1cm}\vspace*{2cm}
\resizebox{14cm}{10cm}
                {\includegraphics{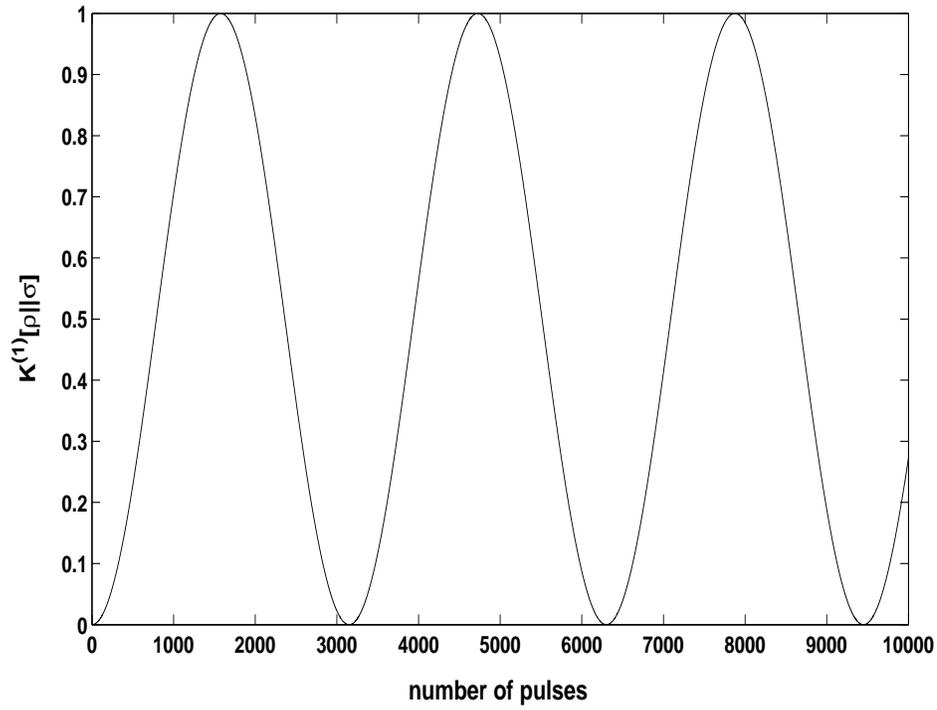}}
\caption{Linear quantum divergence $K^{(1)}_{KL}$ versus the number of the pulses from the external field.
The parameters are: $\epsilon=0.1$; $\Delta_\epsilon=0.001$, $T=\pi$, $\chi=1$.}
\end{figure}

\begin{figure}[p]
\resizebox{14cm}{10cm}
                {\includegraphics{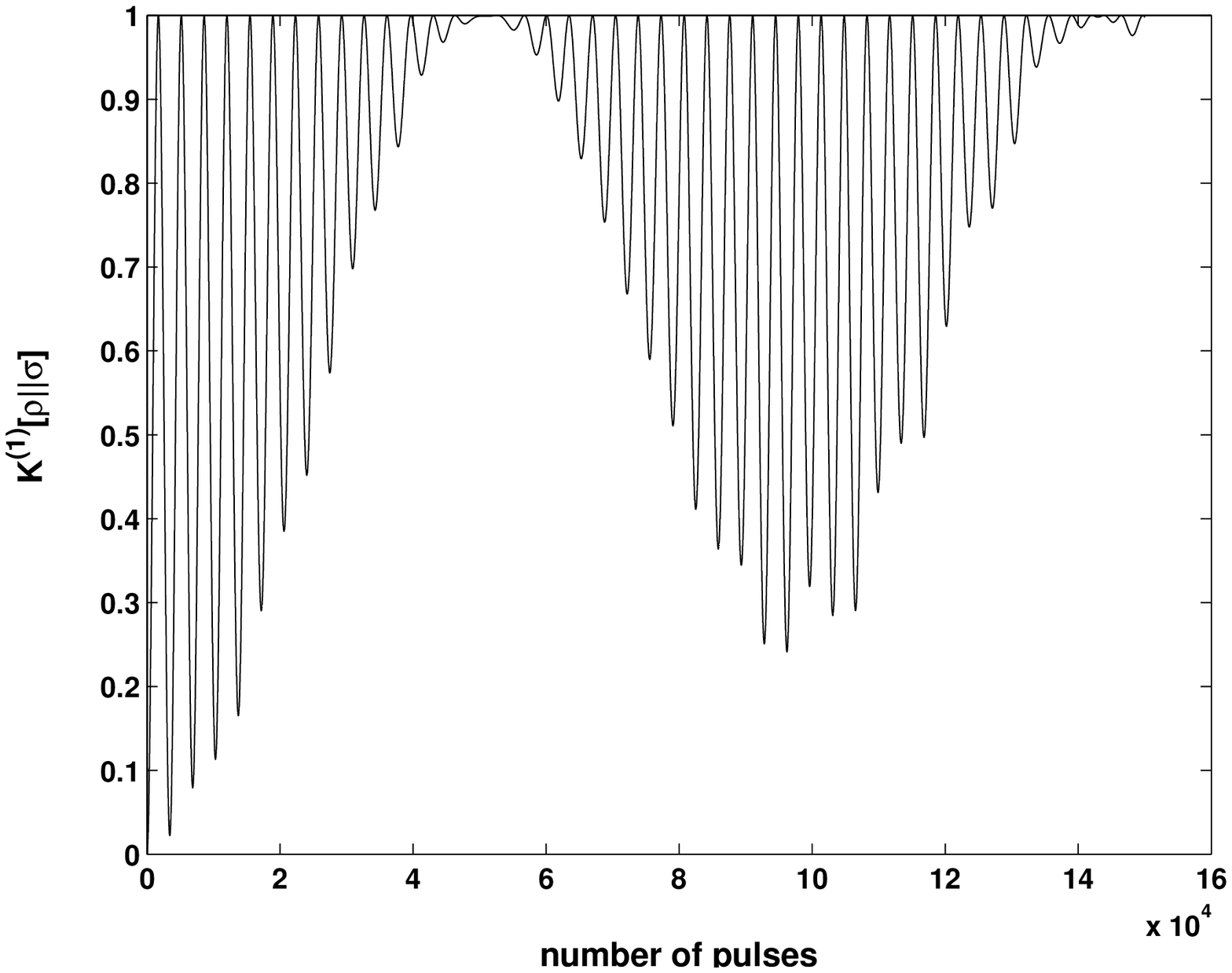}}
\caption{Same as in Fig. 2 but for $\epsilon=0.46$.}
\end{figure}
\begin{figure}[p]
\resizebox{14cm}{7cm}
                {\includegraphics{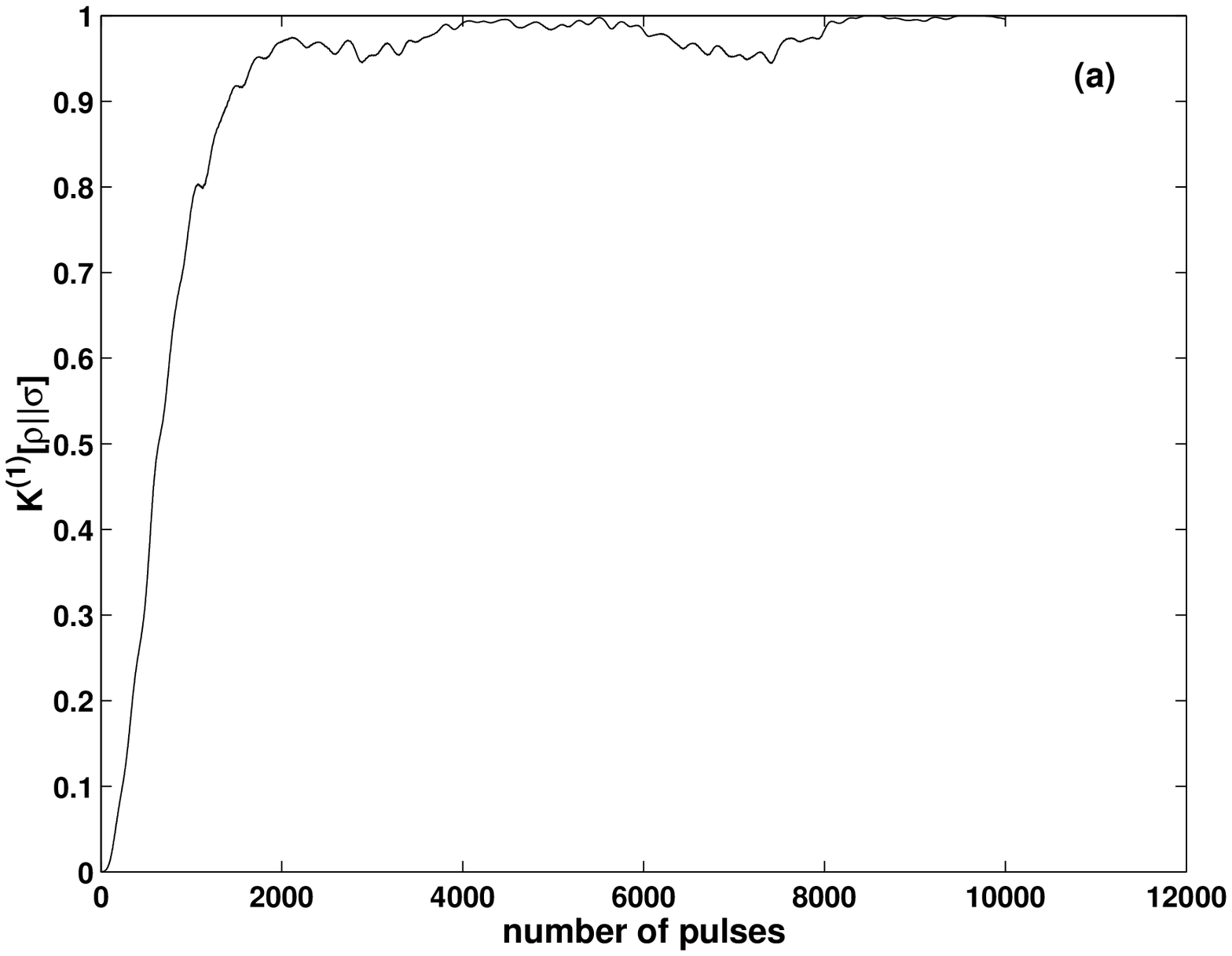}}
\resizebox{14cm}{7cm}
                {\includegraphics{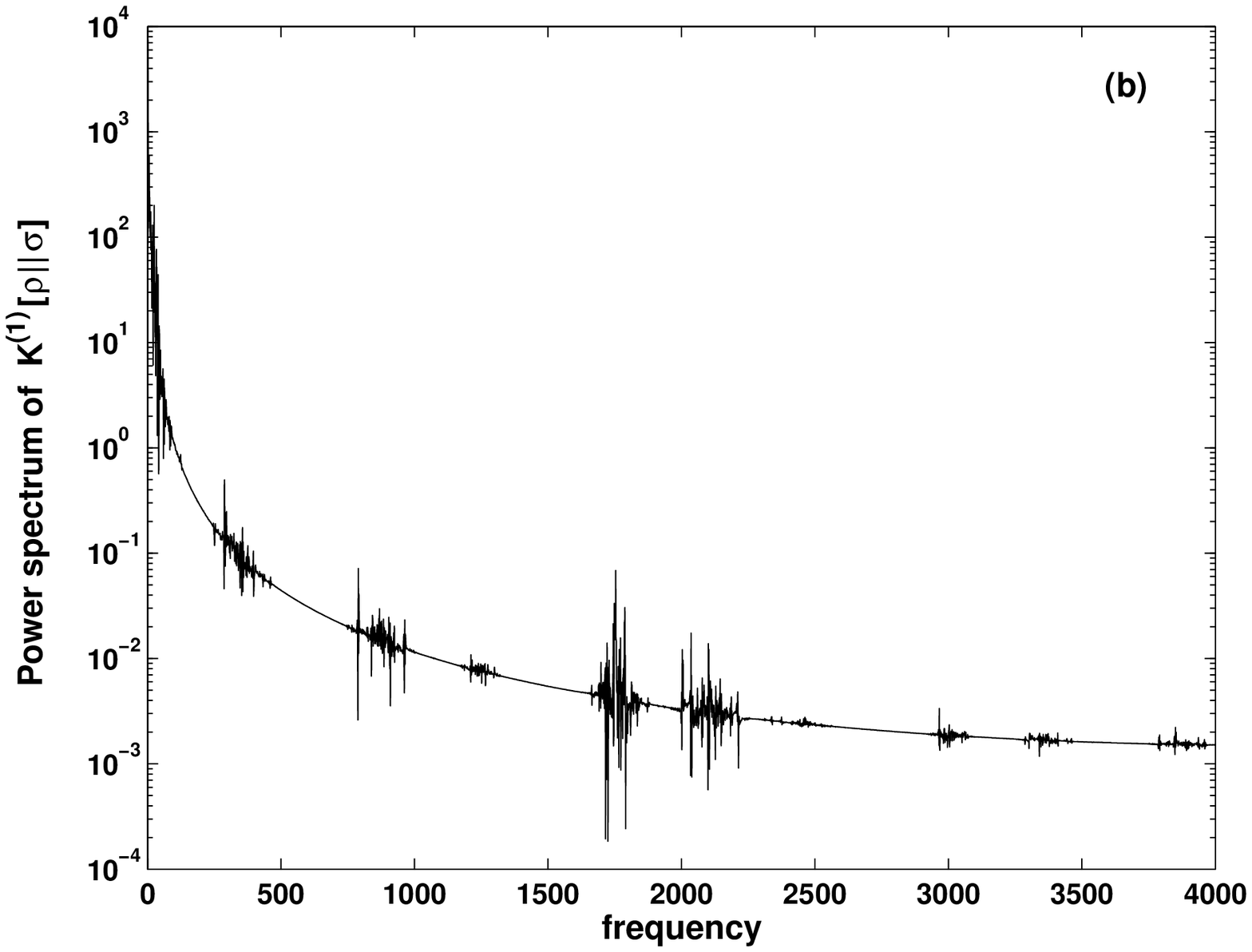}}
\caption{Linear  quantum divergence $K^{(1)}_{KL}$ versus the number of the pulses from the external field -- (a) and the power spectrum of $K^{(1)}_{KL}$ (from 1500 to 10000 pulses) -- (b) for $\epsilon=0.7$; other parameters are the same as in Fig 2.}
\end{figure}
\begin{figure}[p]
\resizebox{14cm}{10cm}
                {\includegraphics{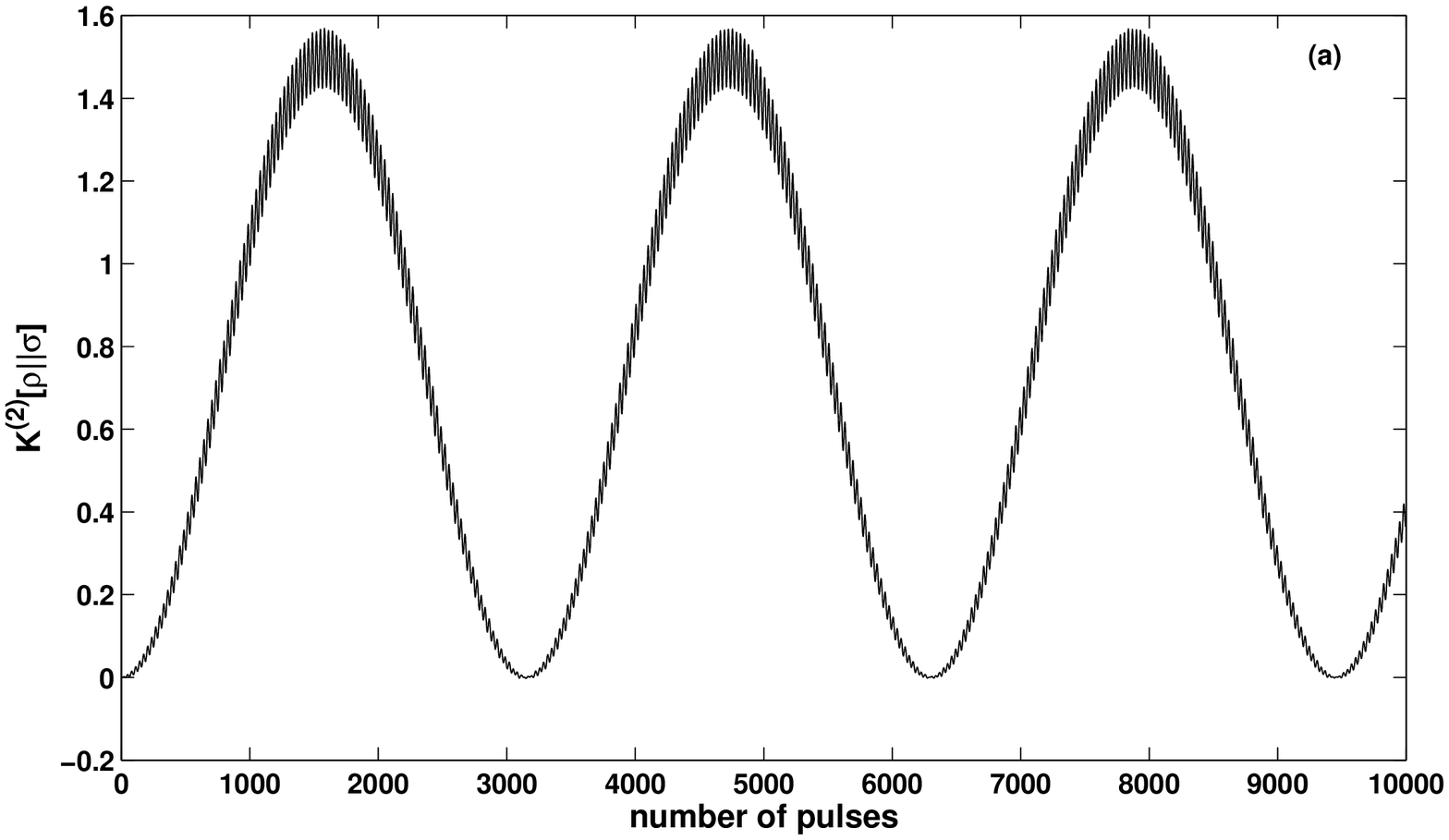}}
\resizebox{14cm}{10cm}
                {\includegraphics{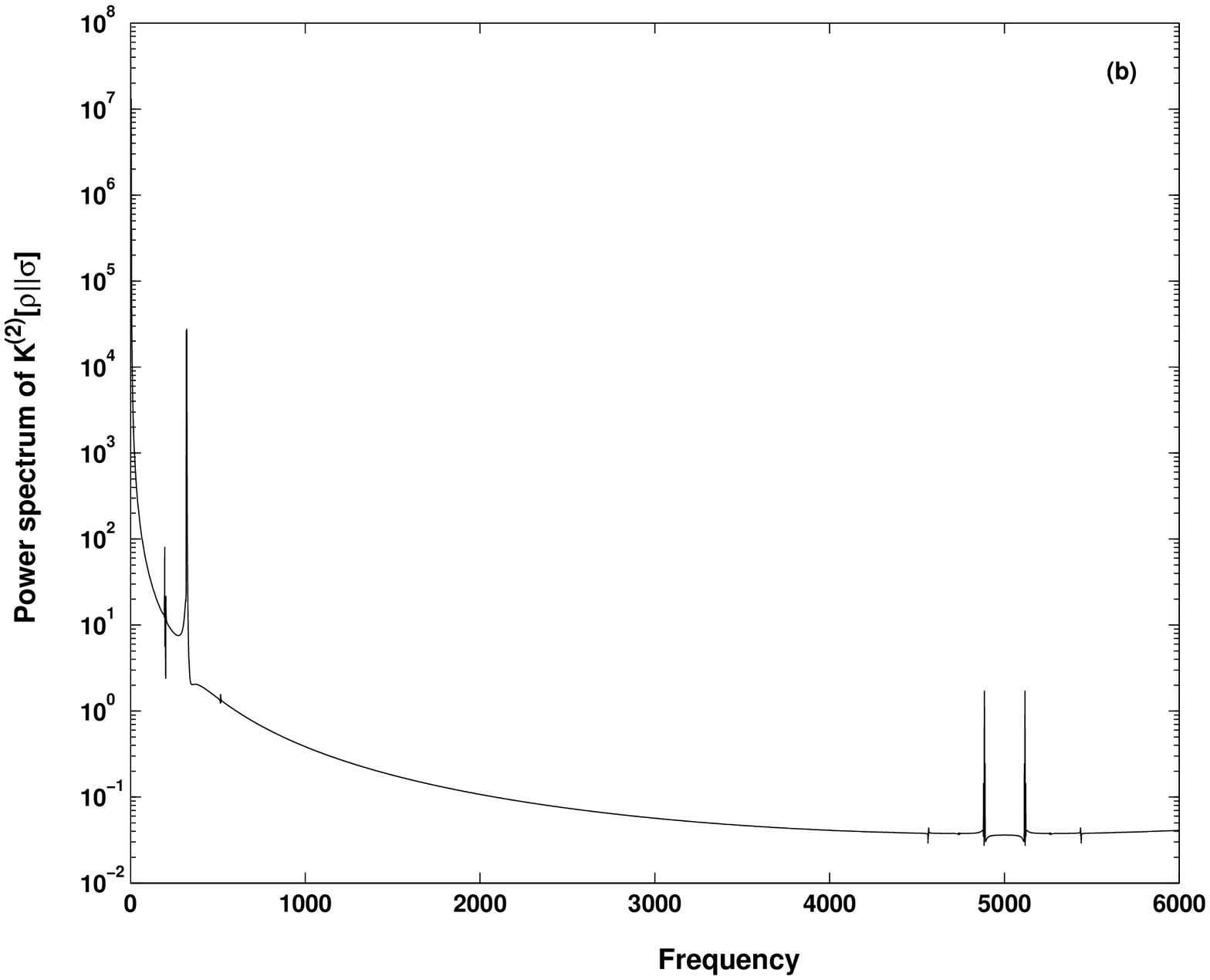}}
\caption{Nonlinear quantum divergence $K^{(2)}_{KL}$ versus the number of the pulses from the external field -- (a) and the power spectrum of $K^{(2)}_{KL}$ -- (b).
for $\epsilon=0.1$; other parameters are the same as in Fig 2.}
\end{figure}
\begin{figure}[p]
\resizebox{14cm}{10cm}
                {\includegraphics{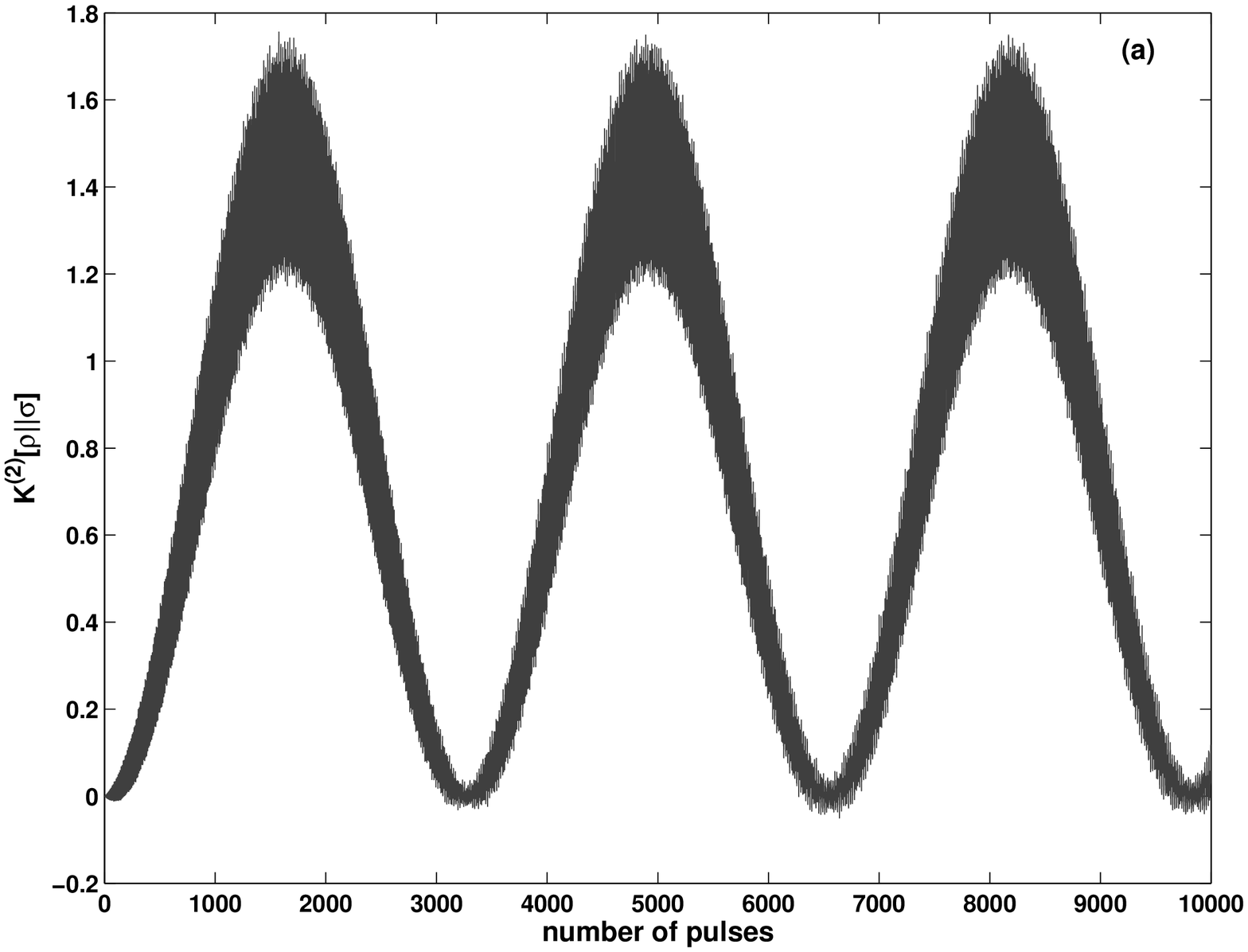}}
\resizebox{14cm}{10cm}
                {\includegraphics{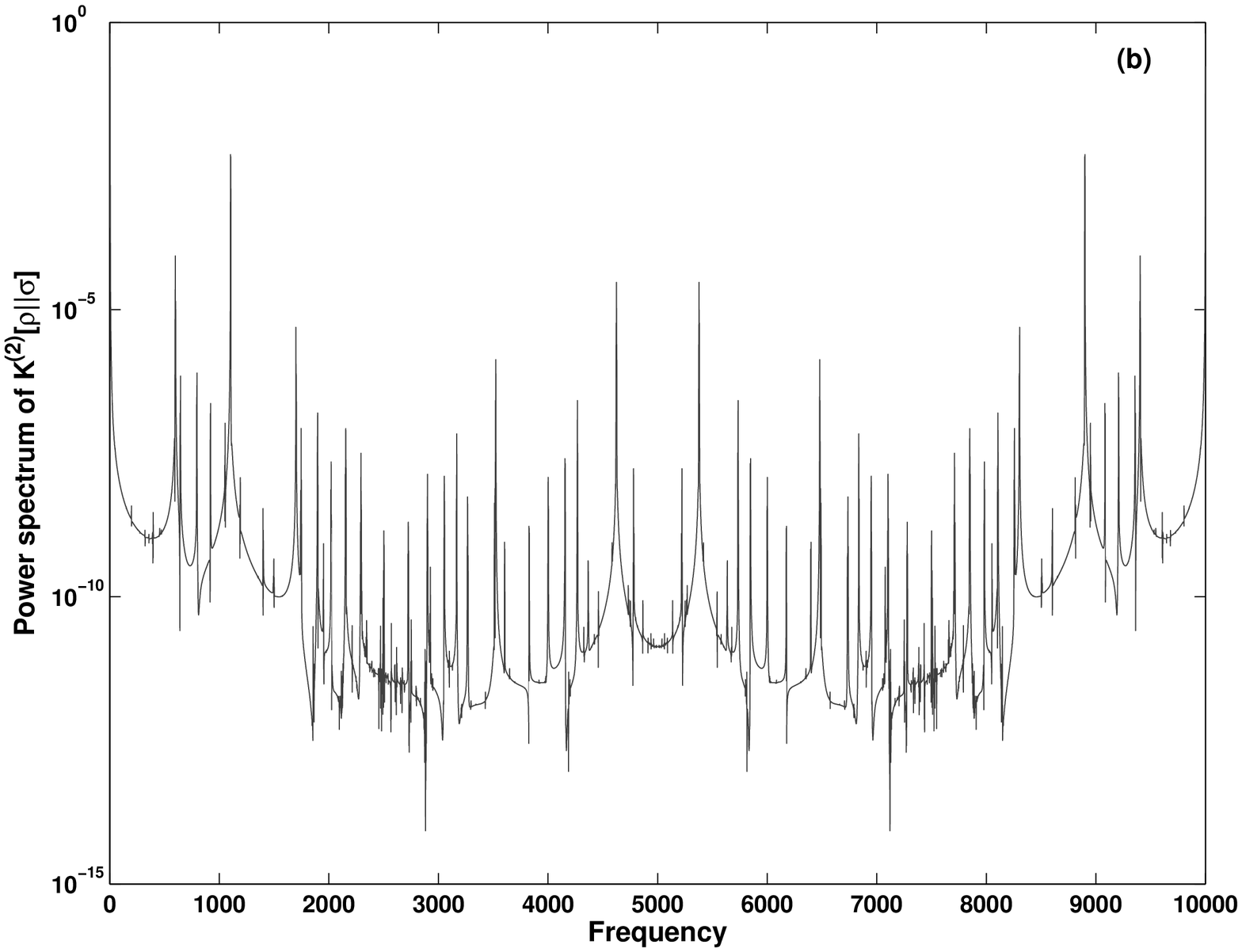}}
\caption{Same as in Fig. 5, but for $\epsilon=0.36$.}
\end{figure}
\begin{figure}[p]
\resizebox{14cm}{10cm}
                {\includegraphics{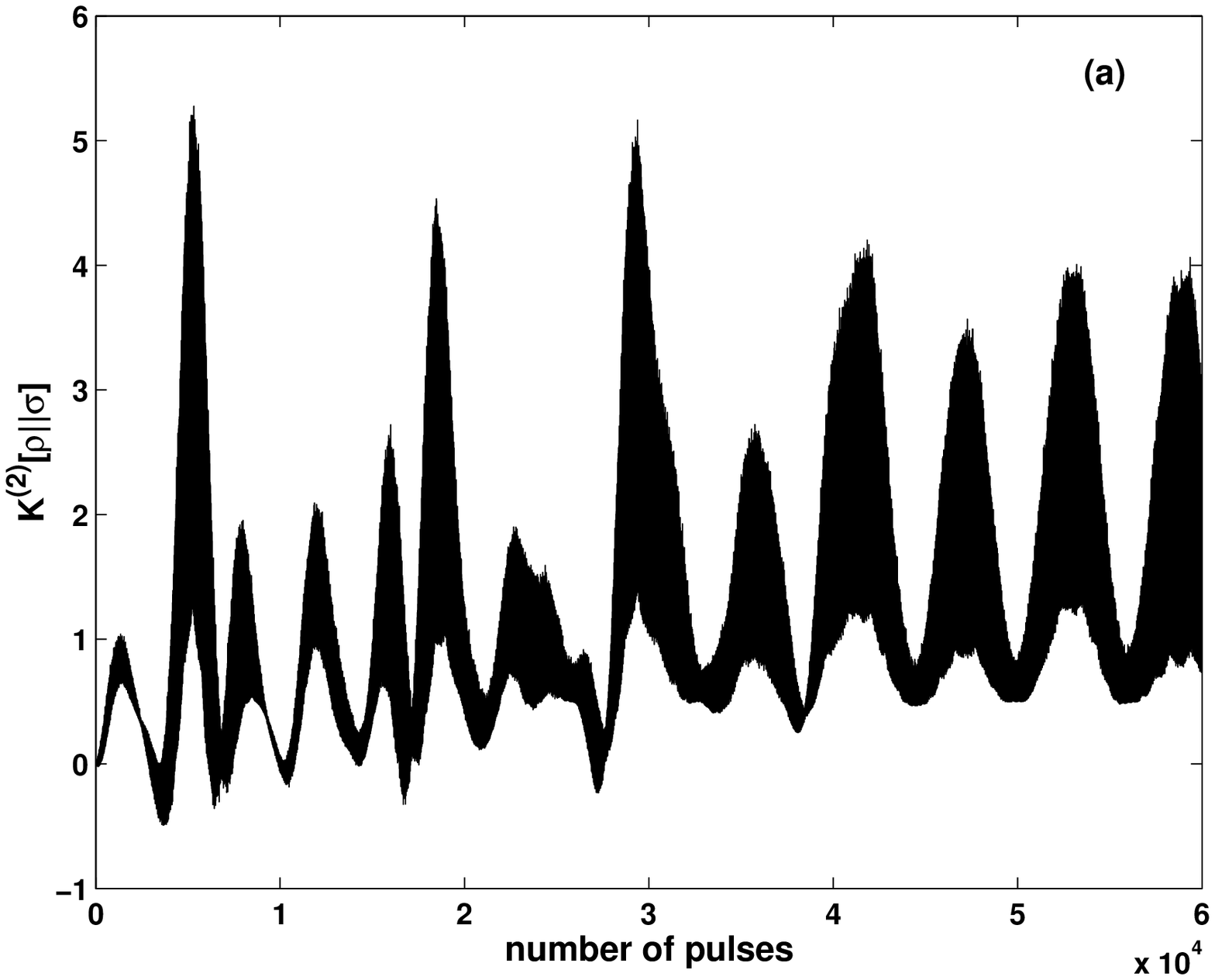}}
\resizebox{16cm}{10cm}
                {\includegraphics{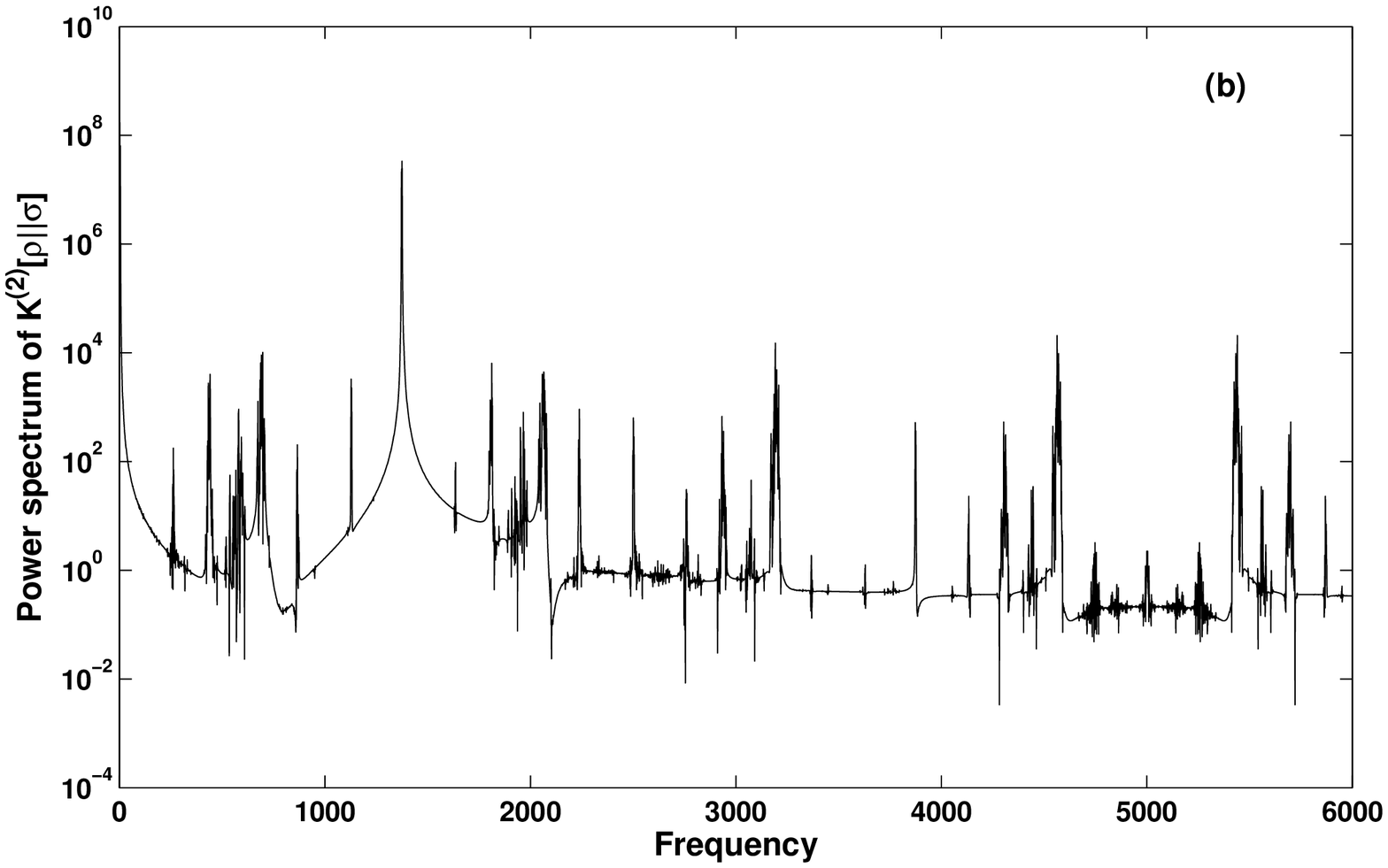}}
\caption{Same as in Fig. 5, but for  $\epsilon=0.46$.}
\end{figure}
\begin{figure}[p]
\resizebox{14cm}{10cm}
                {\includegraphics{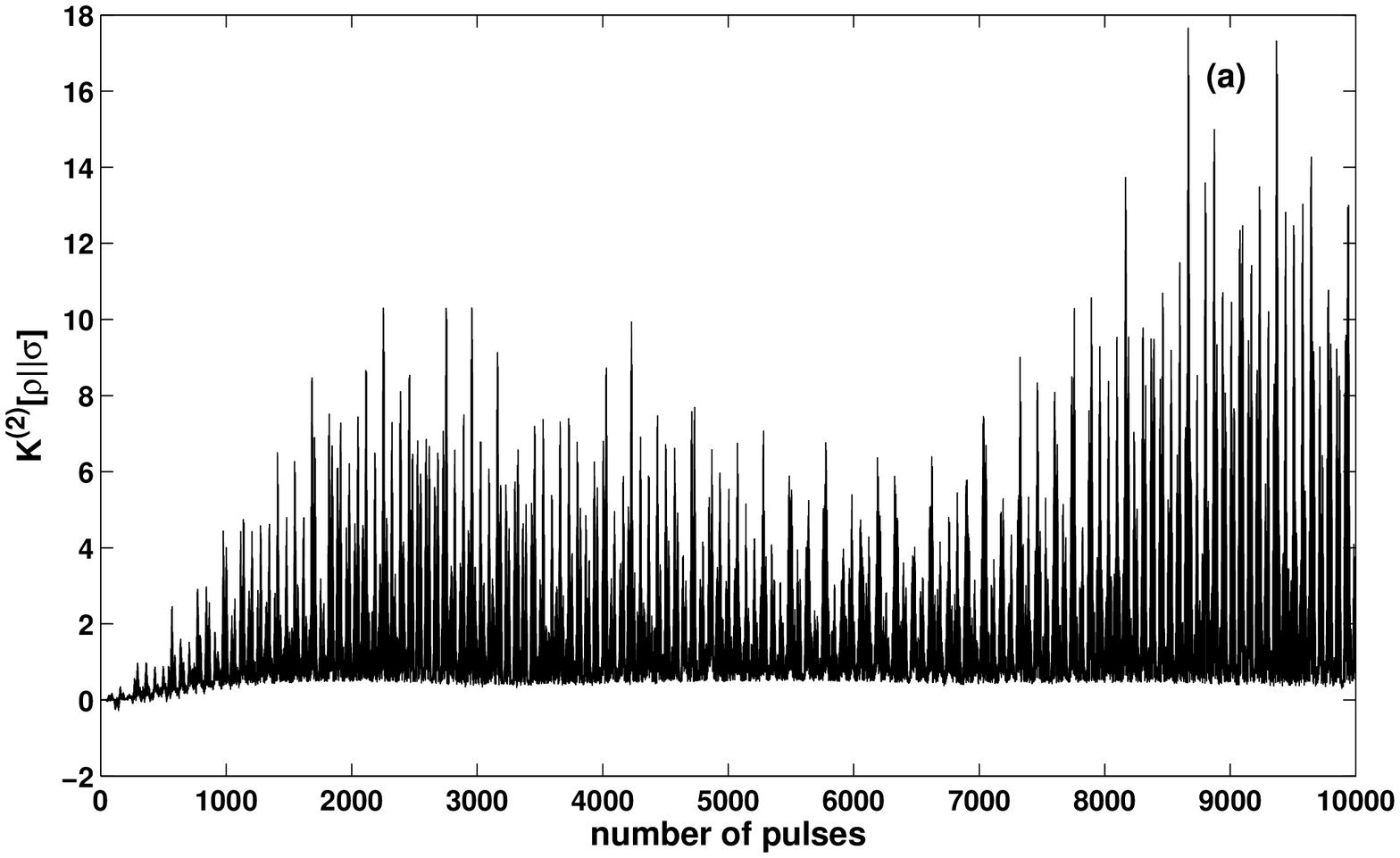}}
\resizebox{14cm}{10cm}
                {\includegraphics{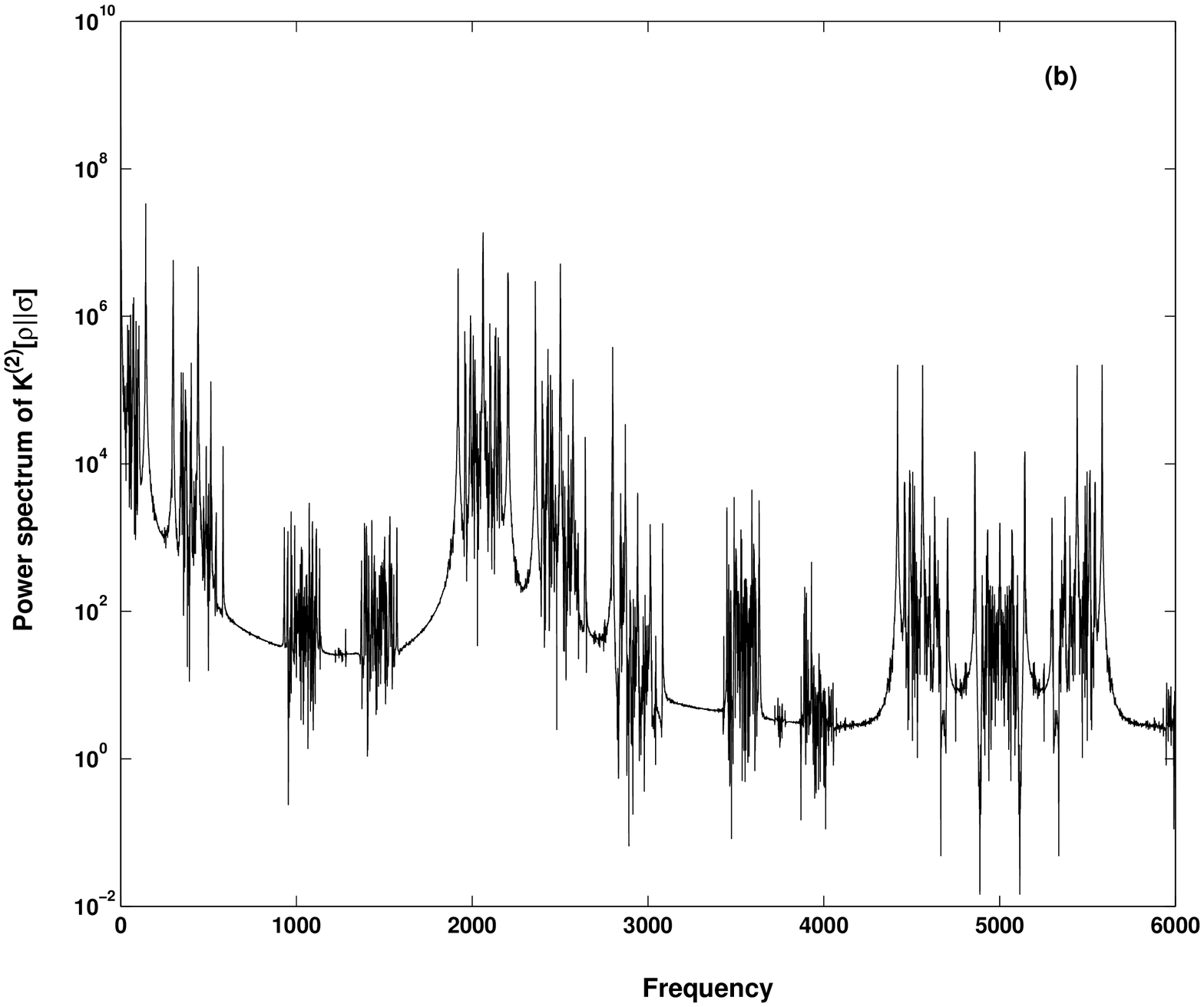}}
\caption{Same as in Fig.5 , but for $\epsilon=0.7$.}
\end{figure}
\end{document}